\documentclass[%
reprint,
superscriptaddress,
 amsmath,amssymb,
 aps,
showpacs,preprintnumbers]{revtex4-1}

\usepackage{graphicx}% Include figure files
\usepackage{dcolumn}% Align table columns on decimal point
\usepackage{bm}% bold math
\usepackage{breqn}
\begin{document}

%\preprint{APS/123-QED}
%\preprint{XXX}

\title{Exact solution of an exciton energy for a monolayer medium}

\author{Abdullah Guvendi}
\author{Ramazan Sahin}
\author{Yusuf Sucu}%
 \email{ysucu@akdeniz.edu.tr}
\affiliation{Department of Physics Faculty of Science Akdeniz University 07058 Turkey}

\date{\today}% It is always \today, today,

\begin{abstract}

We present exact solutions of an energy spectrum of 2-interacting particles in which they seem to be relativistic fermions in 2+1 space-time dimensions. The 2$\times$2 spinor equations of 2-interacting fermions through general central potential were separated covariantly into the relative and center of mass coordinates. First of all, the coupled first order differential equations depending on radial coordinate were derived from 2$\times$2 spinor equations. Then, a second order radial differential equation was obtained and solved for Coulomb interaction potential. We apply our solutions to exciton phenomena for a free-standing monolayer medium. Since we regard exciton as isolated 2-interacting fermions in our model, any other external effect such as substrate was eliminated. Our results show that the obtained binding energies in our model are in agreement with the literature. Moreover, the decay time of an exciton was found out spontaneously in our calculations. 

\end{abstract}

%\pacs{XXX}% PACS, the Physics and Astronomy
\maketitle

%\tableofcontents

\section{Introduction}

Since its invention of monolayer form \cite{graphene_2004}, graphene has attracted noticeable interest due to its extraordinary optical \cite{graphene_optical}, electrical \cite{graphene_electric} and structural \cite{graphene_structure} properties apart from its bulk form of graphite especially. Nowadays, another family of atomically thin materials, transition metal dichalcogenides (TMDs), have came into interest \cite{2D_tmds}.

As direct-gap semiconductors \cite{tmd_direct_gap}, the monolayer members of TMD family ($WS_{2}$, $WSe_{2}$, $MoS_{2}$, $MoSe_{2}$ etc.) have a strong potential in optoelectronic applications \cite{opto_2D}. In semiconductors, promoting of an electron from valance to conduction band leaving a hole behind creates an exciton by photo-excitation which especially finds applications in monolayer TMDs due to higher binding energies than in that of bulk semiconductors \cite{exciton_low_binding, hanbicki_2015} The spatial distance of these two particles is a result of screening effect of surrounding medium through Coulomb interaction \cite{exciton_theory}. These excitons can be used in variety of applications \cite{exciton_app1, exciton_app2}. For example, extending exciton lifetimes in monolayer TMDs give possibility of energy storage in excitonic dark states even at room temperatures \cite{simsek_storage}. 

Although there are much experimental studies focusing on measurement of exciton binding energies for monolayer TMDs \cite{cao_2014, chernikov_2014} and theoretical approaches for prediction of energy spectra of excitonic states \cite{exciton_theory1, exciton_theory2, exciton_theory3}, the obtained results and predictions vary in the range of 0.3 - 1 eV \cite{hanbicki_2015, chernikov_2014, cao_2014}. Some research groups explain the differences by the substrate effects in which the monolayer TMD sample rely on it \cite{binding_substrate}. In addition, triangular lattice model was developed in order to explain non-hydrogen-like behavior of these quasi-particles \cite{triangular_exciton}. In the traditional approach; the exciton is regarded as a single particle where the electron is attracted by the Coulomb potential of an hole where screening effect gives possibility of excitonic states.  

% and its sub-theory like QED, QCD (referans vermelisin)

On the other hand, exploring of relativistic dynamics of interacting particles has been very attractive in Quantum Field Theory (QFT). Although these topics have been studied for a long time, there are still much efforts to explain significant differences between experimental and theoretical results \cite{baryon_spectrum}. For example, relativistic two-body equation was written first by Eddington and Gaunt in 1929 \cite{tale_3_equations}. In addition, a new relativistic Two Body Equation (TBE) was introduced short after \cite{breit_1929} which is relatively straightforward for interacting spin-1/2 particles since it includes approximate Darwin Lagrangian as an interaction term and 2-free Dirac Hamiltonian. This effective Hamiltonian fails either the velocity of particles or distance between particles is very high. Therefore, it covers actually weak coupling approximation. \\

%($\frac{v^{2}}{c^{2}}$) limit in

Another research group obtained a different formulation of relativistic equation \cite{bethe_salpeter_1951} in the Quantum Field Theory (QFT) which satisfies the physics in \cite{breit_1929} up to the non-relativistic regime. Since they found negative solutions for binding energy, their equation was not valid as a bound-state solution of 2-interacting particles due to relative time approach.\\

%In single-time regime

Since we focus on the exciton interaction energy, bound-state equation of relativistic 2-Dirac particles is very important \cite{kemmer_1937,fermi-yang}. However, the interaction potential was written phenomenologically in these studies. A complete equation which is quite similar to \cite{kemmer_1937,fermi-yang}, starting from Quantum Electro-Dynamics (QED) and using Lagrange formalism assumed as a first principal in theoretical physics has been derived. This equation consists of most general electric and magnetic potentials \cite{barut_derivation_1985}. Therefore, its solution gives us a well-known spectra similar to hydrogen-like atoms in all order ($\alpha^6$) even though it could not be solved completely \cite{barut_radial_1985}.\\

%We studied Barut's equation
One can use equations in \cite{barut_derivation_1985} for solving of 2-body problem in (2+1) dimensions due to axial symmetry in Coulomb-like potentials and to obtain especially binding energy of electron-hole coupling (exciton) in monolayer materials. Although evaluating equations in (3+1) space-time is sometimes more complicated \cite{baryon_spectrum,meson_spectroscopy,relativistic_two_fermion_problem,oleg-klaus}, published results studied in (2+1) dimensions \cite{dong_exact_2003,witten_1988} indicate that obtained results in (2+1) dimensions are very close to results obtained in (3+1) dimensions. Therefore, the direct solution of an exciton binding energy would be perfect regardless of surrounding medium. In this current work, exciton is regarded as 2 interacting oppositely charged fermions. Moreover, the solutions were obtained in 2+1 dimensions for a free-standing monolayer medium in order to eliminate any other external effects such as substrate, surrounding medium. For these purposes, two-body problem is exactly solved in 2+1 dimensions, firstly. Then, exciton binding energies and decay times are obtained in terms of spin and energy level of interacting particles as follows.

\section{Two-body Dirac Equation in (2+1) Dimensions}

%Barut's two-body Dirac Equation in (2+1) Dimensions and Fully
We first start from two-body Dirac equation in (3+1) dimensions \cite{barut_derivation_1985} and fully covariant separation of center of mass and relative variable for acquiring energy spectrum of 2-interacting particles.

%\begin{multline}

\begin{dmath}
\left\{ \left[ \gamma _{\mu }^{\left( 1\right) }i\hslash \left( \partial
_{\mu }^{\left( 1\right) }-e_{1}A_{\mu }^{\left( 2\right) }\right)
-m_{1}c\right] \otimes \gamma _{0}^{\left( 2\right) }+\gamma _{0}^{\left(
1\right) }\otimes \left[ \gamma _{\mu }^{\left( 2\right) }i\hslash \left(
\partial _{\mu }^{\left( 2\right) }-e_{2}A_{\mu }^{\left( 1\right) }\right)
-m_{2}c\right] \right\} \phi \left( \mathbf{X}_{1},\mathbf{X}_{2} \right) = 0
\label{Eq1}
\end{dmath}

Since the Dirac matrices have been chosen as $\gamma _{0}=\sigma^{3},\quad\gamma _{1}=i\sigma ^{1}$ and $\quad \gamma _{2}=i\sigma ^{2}$ satisfying Dirac algebra and two-body Dirac spinor is defined by Eq. \ref{Eq_second} as follows;

\begin{dmath}
\phi \left( \textbf{X}_{1},\textbf{X}_{2}\right) =\binom{\binom{D_{1}}{D_{2}}}{\binom{D_{3}}{D_{4}}}
\label{Eq_second}
\end{dmath}

Eq. \ref{Eq1} is reduced to (2+1) dimensions. Thus, we could obtain a coupled equation set from Eq. \ref{Eq1} where $D_{i}=D_{i}(\textbf{r,R},R_{0})$. The explicit form of $D_{i}$ is given by the following expression.

\begin{eqnarray*}
D_{i} = \Psi _{i}\left(r\right) \Phi \left( R\right) e^{-iwR_{0}} , \Phi \left( R\right) = e^{-i\mathbf{k.R}}
\end{eqnarray*}

Here, $D_i$'s where i=1,2,3,4 are the spinor components and we use relative ($r$) and center of mass ($R$) coordinates explicitly. Then, we introduce the (2+1)-dimensional center of mass and relative variables with the following expressions.

\begin{equation}
\begin{split}
R & =\frac{1}{M}\left( m_{1}X_{1}+m_{2}X_{2}\right) \\ 
& r=X_{1}-X_{2} \\
& M=m_{1}+m_{2} \\ 
& \textbf{X}_{1}=(x^{(1)},y^{(1)},t^{(1)}) \\ 
& \textbf{X}_{2}=(x^{(2)},y^{(2)},t^{(2)})
\end{split}
\end{equation}

Here, $m_1$ and $m_2$ are the masses of particles where $X_1$ and $X_2$ are the magnitudes of position vectors of corresponding particles in 2+1 dimensions, respectively. Writing positions of particles in terms of center of mass and relative coordinates

\begin{equation}
\begin{split}
X_{1}=\frac{m_{2}}{M}\left( r+\frac{M}{m_{2}} R\right)\\
X_{2}=\frac{m_{1}}{M}\left( -r+\frac{M}{m1}R\right)
\end{split}
\end{equation}

and partial derivative of $X_1$ and $X_2$ give us the following equations.

\begin{equation}
\begin{split}
\frac{\partial^{(1)} }{\partial X_{\mu }%
}=\frac{\partial }{\partial r_{\mu }}+\frac{m_{1}}{M}\frac{\partial }{%
\partial R_{\mu }}\\
\frac{\partial^{(2)} }{\partial X_{\mu }}=-\frac{\partial
}{\partial r_{\mu }}+\frac{m_{2}}{M}\frac{\partial }{\partial R_{\mu }}\\
\frac{\partial^{(1)} }{\partial X_{\mu }}+\frac{\partial^{(2)} }{\partial X_{\mu }}=\frac{\partial }{\partial R_{\mu }}\\
\frac{\partial }{\partial r_{\mu }}=\frac{1}{M}\left( m_{2}\frac{%
\partial^{(1)} }{\partial X_{\mu }}-m_{1}\frac{\partial^{(2)} }{\partial X_{\mu }}%
\right) 
\label{eq_5}
\end{split}
\end{equation}

where $\mu =0,1,2$  and

%%%%%%%%%%%%%%%%%%%%%%%%%%%%%%%%%%%%%
\quad  $\partial _{0}=\partial _{t}/c,\quad \partial _{1}=\partial _{x},\quad\partial
_{2}=\partial _{y}$. \\
%%%%%%%%%%%%%%%%%%%%%%%%%%%%%%%%%%%%%

For central interaction; the components of vector potential are taken as follows $A_{0}=V\left( \mathbf{x}_{1}-\mathbf{x}_{2}\right),\quad A_{1}=0,\quad A_{2}=0$.

\section{Derivation of Radial Equations}

We can apply a method of separation of variables in terms of relative and center of mass coordinates provided that a central potential is taken into account; $V\equiv V\left( \textbf{r}\right)$.

In our calculations, we assume that the center of mass does not carry a momentum because we take k as $\bf{k}$=0. Therefore, one can write set of equation in terms of relative coordinates.

In the second part of calculations, we reorganized the coupled equation set by inserting Eq. \ref{eq_5} and definition of $D_{i}$, we obtained the following equations.

\begin{multline}
\left( \frac{w}{c}-\frac{Mc}{\hbar} -iV_{t}(r)\right) \Psi
_{1}+\left( \partial _{r_{1}}-i\partial _{r_{2}}\right) \Psi _{2} \\ -\left( \partial _{r_{1}}-i\partial _{r_{2}}\right) \Psi _{3}=0,\label{equation4}
\end{multline}

\begin{multline}
\left( -\frac{w}{c}+\frac{\triangle mc}{\hslash} +iV_{t}(r)\right) \Psi _{2}+\left( \partial _{r_{1}}+i\partial _{r_{2}}\right) \Psi _{1}  \\ +\left( \partial _{r_{1}}-i\partial _{r_{2}}\right) \Psi _{4}=0,\label{equation5}
\end{multline}

\begin{multline}
\left( -\frac{w}{c}-\frac{\triangle mc}{\hslash} +iV_{t}(r)\right) \Psi_{3}-\left( \partial _{r_{1}}-i\partial _{r_{2}}\right) \Psi _{4} \\
-\left( \partial _{r_{1}}+i\partial _{r_{2}}\right) \Psi _{1}=0,\label{equation6}
\end{multline}

\begin{multline}
\left( \frac{w}{c}+\frac{Mc}{\hbar} -iV_{t}(r)\right) \Psi
_{4}-\left( \partial _{r_{1}}+i\partial _{r_{2}}\right) \Psi _{3}
 \\ +\left( \partial _{r_{1}}+i\partial _{r_{2}}\right) \Psi _{2}=0\label{equation7}
\end{multline}

In order to write the coupled equation set in an explicit form, we use both

\begin{eqnarray}
 \left( \partial _{r_{1}}-i\partial _{r_{2}}\right) = e^{-i\Phi
}\left( -\frac{i}{r}\partial _{\Phi }+\partial _{r}\right)
\end{eqnarray}

 and

 \begin{eqnarray}
\left( \partial _{r_{1}}+i\partial _{r_{2}}\right)=e^{i\Phi }\left( \frac{i%
}{r}\partial _{\Phi }+\partial _{r}\right)
\end{eqnarray}

definitions and obtained the following explicit equation set in polar coordinates. In addition, we multiply the Eq. \ref{equation4} by $e^{i\Phi }$ and the Eq. \ref{equation7} by $e^{-i\Phi}$ on the left side. For the Eq. \ref{equation5} and Eq. \ref{equation6}, the phase terms ($e^{i\Phi }$ and $e^{-i\Phi}$) are required to be positioned in front of the operators paving the way that an additional term is included in front of the corresponding operator. By conducting these, we end up with the following equations;

%\begin{multline}
%\left( \frac{w}{c}-\frac{Mc}{\hbar} -iV_{t}(r)\right) \Psi _{1}\left(
%r,\Phi \right) \\ + e^{-i\Phi }\left( -\frac{i}{r}\partial _{\Phi
%}+\partial _{r}\right)\Psi _{2}\left( r,\Phi \right)
%\\ - e^{-i\Phi }\left( -\frac{i}{r}\partial _{\Phi }+\partial _{r}\right)\Psi _{3}\left( r,\Phi
%\right) =0\label{eq_set1}
%\end{multline}

%\begin{multline}
%\left( -\frac{w}{c}+\frac{\triangle mc}{\hslash} +iV_{t}(r)\right) \Psi
%_{2}\left( r,\Phi \right) \\ + e^{i\Phi }\left( \frac{i}{r}\partial
%_{\Phi }+\partial _{r}\right)\Psi _{1}\left( r,\Phi \right)
%\\ + e^{-i\Phi }\left( -\frac{i}{r}%
%\partial _{\Phi }+\partial _{r}\right)\Psi _{4}\left( r,\Phi
%\right) =0\label{eq_set2}
%\end{multline}

%\begin{multline}
%\left( -\frac{w}{c}-\frac{\triangle mc}{\hslash} +iV_{t}(r)\right) \Psi
%_{3}\left( r,\Phi \right) \\ - e^{-i\Phi }\left( -\frac{i}{r}\partial
%_{\Phi }+\partial _{r}\right)\Psi _{4}\left( r,\Phi \right)
%\\ - e^{i\Phi }\left( \frac{i}{r}%
%\partial _{\Phi }+\partial _{r}\right)\Psi _{1}\left( r,\Phi
%\right) =0,\label{eq_set3}
%\end{multline}

%\begin{multline}
%\left( \frac{w}{c}+\frac{Mc}{\hbar} -iV_{t}(r)\right) \Psi _{4}\left(
%r,\Phi \right) \\ - e^{i\Phi }\left( \frac{i}{r}\partial _{\Phi
%}+\partial _{r}\right)\Psi _{3}\left( r,\Phi \right)
%\\ + e^{i\Phi }\left( \frac{i}{r}%
%\partial _{\Phi }+\partial _{r}\right)\Psi _{2}\left( r,\Phi
%\right) =0\label{eq_set4}
%\end{multline}

\begin{multline}
\left( \frac{w}{c}-\frac{Mc}{\hbar} -iV_{t}(r)\right) \Psi _{1}\left(r,\Phi \right) e^{i\Phi } \\ + \left( -\frac{i}{r}\partial _{\Phi }+\partial_{r}\right) \Psi _{2}\left( r,\Phi \right)
\\ -\left(-\frac{i}{r}\partial _{\Phi }+\partial _{r}\right) \Psi _{3}\left(r,\Phi \right) =0\quad \label{equation20}
\end{multline}

\begin{multline}
\left( -\frac{w}{c}+\frac{\triangle mc}{\hslash} +iV_{t}(r)\right) \Psi_{2}\left( r,\Phi \right) \\ + \left( \frac{i}{r}\partial _{\Phi }+\frac{1}{r}+\partial
_{r}\right) \Psi _{1}\left( r,\Phi \right) e^{i\Phi }
 \\ + \left( -\frac{i}{r}\partial _{\Phi }+\frac{1}{r}+\partial _{r}\right) \Psi
_{4}\left( r,\Phi \right) e^{-i\Phi }=0\label{equation21}
\end{multline}

\begin{multline}
\left( -\frac{w}{c}-\frac{\triangle mc}{\hslash} +iV_{t}(r)\right) \Psi_{3}\left( r,\Phi \right) \\ - \left( -\frac{i}{r}\partial _{\Phi }
+\frac{1}{r}+\partial_{r}\right) \Psi _{4}\left( r,\Phi \right) e^{-i\Phi }
 \\ - \left( \frac{i}{r}\partial _{\Phi }+\frac{1}{r}+\partial _{r}\right) \Psi
_{1}\left( r,\Phi \right) e^{i\Phi }=0\label{equation22}
\end{multline}

\begin{multline}
\left( \frac{w}{c}+\frac{Mc}{\hbar} -iV_{t}(r)\right) \Psi _{4}\left(
r,\Phi \right) e^{-i\Phi } \\ - \left( \frac{i}{r}\partial _{\Phi }+\partial_{r}\right) \Psi _{3}\left( r,\Phi \right)
 \\ +\left( \frac{i}{r}\partial _{\Phi }+\partial _{r}\right) \Psi _{2}\left(r,\Phi \right) =0\label{equation23}
\end{multline}

where $\Psi _{g}\left( r,\Phi \right) =\Psi _{g}\left( r\right)
e^{ij_{g}\Phi }$ \ $\left( g=1,2,3,4\right). $\\

Since the $\Phi$ is a cyclic coordinate, we derive a coupled first order 4-differential equations by using this property. Then, multiplying the Eq. \ref{equation20} by $\left( \frac{w}{c}+\frac{Mc}{\hbar} -iV_{t}(r)\right) $, the Eq. \ref{equation23} by $\left( \frac{w}{c}-\frac{Mc}{\hbar} -iV_{t}(r)\right) $, the Eq. \ref{equation21} by $\left( -\frac{w}{c}-\frac{\triangle mc}{\hslash} +iV_{t}(r)\right) $ and the Eq. \ref{equation22} by $\left( -\frac{w}{c}+\frac{\triangle mc}{\hslash} +iV_{t}(r)\right) $ on the left side of equations cause in a new definition set indicated as follows.

%\begin{eqnarray*}
%\left( \Psi _{2}\left( r\right) -\Psi _{3}\left( r\right) \right) = \Psi
%_{0}\left( r\right)\\
%\left( \Psi _{1}\left( r\right) e^{i\Phi }+\Psi _{4}\left(
%r\right) e^{-i\Phi }\right) = \Psi ^{+}\left( r\right),\\
%\left( \Psi_{1}\left( r\right) e^{i\Phi }-\Psi _{4}\left( r\right) e^{-i\Phi }\right)
%= \Psi ^{-}\left( r\right)
%\end{eqnarray*}

\section{Derivation of Complete $2^{nd}$ Order Equation}

By pursuing above mathematical procedures and substituting $\Psi _{g}\left( r,\Phi \right)$, we obtained the following equations for general central potentials.

\begin{multline}
\left( U(r)^{2}-M_{1}^{2}\right) \Psi ^{+}(r) \\ +\ 2\left( M_{1}%
\text{ }\frac{j}{r}+U(r)\frac{d}{dr}\right) \Psi _{0}\left( r\right) = 0
\label{equation25}
\end{multline}

\begin{multline}
\left( U(r)^{2}-M_{1}^{2}\right) \Psi ^{-}(r) \\ + \ 2\left(
U(r)\text{ }\frac{j}{r}+M_{1}\frac{d}{dr}\right) \Psi _{0}\left( r\right) = 0
\label{equation26}
\end{multline}

\begin{multline}
\left( U(r)^{2}-M_{2}^{2}\right) \Psi _{0}\left( r\right)
\\ +2\left( M_{2}\text{ }\frac{j}{r}\right) \Psi ^{-}(r)-2M_{2}\frac{d}{dr}\Psi ^{+}(r) = 0
\end{multline}

\begin{multline}
\left( U(r)^{2}-M_{2}^{2}\right) \Psi _{0}\left( r\right)
\\ +2\left( U(r)\text{ }\frac{j}{r}\right) \Psi ^{-}(r)-2U(r)\frac{d}{dr}\Psi ^{+}(r) = 0
\label{aaa}
\end{multline}

where $M_{1}=\frac{Mc}{\hslash} ,$ $M_{2}=$, $\frac{\triangle mc}{\hslash} $  and $U(r)=\left( \frac{w}{c}-iV_{t}(r)\right).$ From Eq. \ref{equation25} and Eq. \ref{equation26}, we rewrite $\Psi^{+}(r)$ and $\Psi^{-}(r)$ in terms of $\Psi _{0}\left( r\right) $ .

\begin{eqnarray*}
\Psi ^{+}(r)&=&-\frac{2}{\left(
U(r)^{2}-M_{1}^{2}\right) }\left( M_{1}\text{ }\frac{j}{r}+U(r)\frac{d}{dr}\right) \Psi
_{0}\left( r\right) , \\
\Psi ^{-}(r)&=&-\frac{2}{\left(
U(r)^{2}-M_{1}^{2}\right) }\left( U(r)\text{ }\frac{j}{r}r+M_{1}\frac{d}{dr}\right) \Psi
_{0}\left( r\right)
\end{eqnarray*}

Inserting these expressions into the Eq. \ref{aaa}, we obtain a radial second order differential equation which contains general central potential.

%\begin{dmath}
%\frac{d^{2}}{dr^{2}}\Psi _{0}\left( r\right) +\left[ \frac{U^{^{\prime }}(r)}{U(r)}-\frac{2U(r)U^{^{\prime }}(r)}{\left( U(r)^{2}-M_{1}^{2}\right) }\right]
%\frac{d}{dr}\Psi _{0}\left( r\right) \\ + \left[ \frac{\left(U(r)^{2}-M_{1}^{2}\right) \left( U(r)^{2}-M_{2}^{2}\right) }{4U(r)^{2}}-\frac{j^{2}}{r^{2}}-\frac{2jU^{^{\prime }}(r)M_{1}}{\left( U(r)^{2}-M_{1}^{2}\right) } -\frac{jM_{1}}{U(r)r^{2}}\right] \Psi _{0}\left( r\right) = 0 
%\end{dmath}

\section{Solution for a Coulomb Interaction}

Since the exciton is composed of an electron and hole, the interaction between them can be written in terms of Coulomb potential. In addition, these particles (an electron and a hole) posses very small masses. Therefore, we can neglect the $M_{1}^{2}$ and $M_{2}^{2}$ in our equations.

\begin{multline}
\frac{d^{2}}{dr^{2}}\Psi _{0}\left( r\right) -\left[ \frac{%
U^{^{\prime }}(r)}{U(r)}\right] \frac{d}{dr}\Psi _{0}\left( r\right) \\ + \left[
\frac{U(r)^{2}}{4}-\frac{j^{2}}{r^{2}}-\frac{2jU^{^{\prime }}(r)M_{1}}{U(r)^{2}}%
-\frac{jM_{1}}{U(r)r^{2}}\right] \Psi _{0}\left( r\right) = 0\label{general_equ}
\end{multline}

 Here we defined a new dimensionless variable; $z =\frac{iwr}{2\alpha c}$ for $V_{t}(r)=-2\frac{\alpha }{r}$. Then, the solution of simplified Eq. \ref{general_equ} becomes a HeunC$(\alpha_{1},\beta ,\gamma ,\delta ,\eta$ ,z) function. In this function, frequency relation is found by using $\delta = -(n+\frac{\beta +\gamma +2}{2})\alpha_{1}$ expression. The following parameters;

\begin{eqnarray*}
\alpha_{1}&=&2\alpha,\\\beta &=&-2\sqrt{\alpha ^{2}+j^{2}},\\
\gamma& =&\frac{2}{(\ w)}\sqrt{-4iNc^{2}j\alpha +(\ w)^{2}},\\
\delta&=&2\alpha ^{2}
\end{eqnarray*}
\begin{eqnarray*}
2\alpha ^{2}+2\left(
n-\sqrt{\alpha ^{2}-j^{2}}+\frac{\sqrt{-4iN\alpha c^{2}j+w^{2}}}{w}\right)
\alpha =0
\end{eqnarray*}

yields

\begin{dmath}
w_{\left( n,j\right) }=\pm 2\sqrt{iNc^{2}\alpha j} \left[2\sqrt{\alpha ^{2}-j^{2}}\alpha +2\sqrt{\alpha^{2}-j^{2}}n-2\alpha ^{2} - 2\alpha n-j^{2}-n^{2}+2\sqrt{\alpha ^{2}-j^{2}}-2\alpha -2n \right]^{-1/2}
\end{dmath}

where $N=\frac{Mc}{\hbar} ,$  $M=m_{1}+m_{2}$.

The explicit expression of $w_{\left(n,j\right)}$ allows us to obtain interaction energy of an exciton. This expression covers actually all the energy levels (n) where the particles might be and spin value of interacting particles. For clarity and quantification, a detailed table (Tab. \ref{table1}) is composed of these variables.

\begin{table}[h]\footnotesize
  \caption{Interaction energies for an exciton}
\begin{tabular}{ | c | c | c | c |}
\hline 
    n     &       j       &      E (eV)        & Decay Time (ps)  \\ \hline
    1     &   -1/2        & (0.214 + i 0.214)  & 0.01             \\ 
          &    1/2        & (0.214 - i 0.214)  &                  \\ \hline
          &   -1/2        & (0.105 + i 0.105)  & 0.02             \\  
    2     &    1/2        & (0.105 - i 0.105)  &                  \\ 
					&    3/2        & (0.370 - i 0.370)  & 0.005            \\ \hline
\end{tabular}
\label{table1}
\end{table}

Our calculations are very close to results found in literature \cite{chernikov_2014}. Since our model exclude any other environmental effects such as substrate or crystal structure, we found a little bit lower value for an exciton binding energy. The experimental measurements of an exciton binding energy are based on the photo-excitation of an electron from a valance to a conduction band leaving a hole behind. While the interaction between an electron and a hole is totally attractive, the screening effect allows an exciton to be created. Moreover, we think that the origin of the differences in experimental results of exciton binding energy is due to screening effect depending on the substrate or the crystal structure. Therefore, experimental measurements include this screening energy. Another interesting result of our calculations is that the interaction energy of 2-fermions possesses imaginary part. By using this imaginary part, one can calculate the decay time ($\tau$) of an exciton. Based on the definition in $\tau \propto \frac{1}{w_{\left( n,j\right) }}$, we found relaxation time as 0.01 ps for n=1 condition.

\section{CONCLUSION}

In this work, we found a general definition of interaction energy for 2-interacting fermions. First, we had examined two Dirac particles interacting with their central potential before we derived two-body Dirac equation in (2+1) space-time geometry. For solution of this equation, we separated center of mass and relative coordinates by using explicit form of the equation. Then, we obtained $1^{st}$ order coupled radial differential equation set. Since the masses ($M_1$ and $M_2$) of an electron and a hole are very small, we neglected $M_1^{2}$ and $M_2^{2}$ values in our equations. The solution of 2-interacting fermions in our model indicates that one can obtain general definition for an interaction energy including quantum numbers. We apply our model to an exciton and found its binding energy which is very close to results found in literature. Moreover, since we work in 2+1 dimensions, our results do not include any other effects for a monolayer medium. Finally, we calculated the decay time of an exciton ($\sim$ ps) directly because the interaction energy in our calculations possesses imaginary part.

\begin{acknowledgments}

The authors thank Nuri Unal, Ganim Gecim and Semra Gurtas for usefull discussions.  

\end{acknowledgments}

% for the sample file to show the different styles of references, b ut authors
%\nocite{*}

%\bibliography{mylibrary}% Produces the bibliography via BibTeX.

\begin{thebibliography}{33}%
\makeatletter
\providecommand \@ifxundefined [1]{%
 \@ifx{#1\undefined}
}%
\providecommand \@ifnum [1]{%
 \ifnum #1\expandafter \@firstoftwo
 \else \expandafter \@secondoftwo
 \fi
}%
\providecommand \@ifx [1]{%
 \ifx #1\expandafter \@firstoftwo
 \else \expandafter \@secondoftwo
 \fi
}%
\providecommand \natexlab [1]{#1}%
\providecommand \enquote  [1]{``#1''}%
\providecommand \bibnamefont  [1]{#1}%
\providecommand \bibfnamefont [1]{#1}%
\providecommand \citenamefont [1]{#1}%
\providecommand \href@noop [0]{\@secondoftwo}%
\providecommand \href [0]{\begingroup \@sanitize@url \@href}%
\providecommand \@href[1]{\@@startlink{#1}\@@href}%
\providecommand \@@href[1]{\endgroup#1\@@endlink}%
\providecommand \@sanitize@url [0]{\catcode `\\12\catcode `\$12\catcode
  `\&12\catcode `\#12\catcode `\^12\catcode `\_12\catcode `\%12\relax}%
\providecommand \@@startlink[1]{}%
\providecommand \@@endlink[0]{}%
\providecommand \url  [0]{\begingroup\@sanitize@url \@url }%
\providecommand \@url [1]{\endgroup\@href {#1}{\urlprefix }}%
\providecommand \urlprefix  [0]{URL }%
\providecommand \Eprint [0]{\href }%
\providecommand \doibase [0]{http://dx.doi.org/}%
\providecommand \selectlanguage [0]{\@gobble}%
\providecommand \bibinfo  [0]{\@secondoftwo}%
\providecommand \bibfield  [0]{\@secondoftwo}%
\providecommand \translation [1]{[#1]}%
\providecommand \BibitemOpen [0]{}%
\providecommand \bibitemStop [0]{}%
\providecommand \bibitemNoStop [0]{.\EOS\space}%
\providecommand \EOS [0]{\spacefactor3000\relax}%
\providecommand \BibitemShut  [1]{\csname bibitem#1\endcsname}%
\let\auto@bib@innerbib\@empty
%</preamble>
\bibitem [{\citenamefont {Geim}\ and\ \citenamefont
  {Novoselov}(2007)}]{graphene_2004}%
  \BibitemOpen
  \bibfield  {author} {\bibinfo {author} {\bibfnamefont {A.~K.}\ \bibnamefont
  {Geim}}\ and\ \bibinfo {author} {\bibfnamefont {K.~S.}\ \bibnamefont
  {Novoselov}},\ }\href@noop {} {\bibfield  {journal} {\bibinfo  {journal}
  {Nature Materials}\ }\textbf {\bibinfo {volume} {6}},\ \bibinfo {pages} {183}
  (\bibinfo {year} {2007})}\BibitemShut {NoStop}%
\bibitem [{\citenamefont {Bonaccorso}\ \emph {et~al.}(2010)\citenamefont
  {Bonaccorso}, \citenamefont {Sun}, \citenamefont {Hasan},\ and\ \citenamefont
  {Ferrari}}]{graphene_optical}%
  \BibitemOpen
  \bibfield  {author} {\bibinfo {author} {\bibfnamefont {F.}~\bibnamefont
  {Bonaccorso}}, \bibinfo {author} {\bibfnamefont {Z.}~\bibnamefont {Sun}},
  \bibinfo {author} {\bibfnamefont {T.}~\bibnamefont {Hasan}}, \ and\ \bibinfo
  {author} {\bibfnamefont {A.~C.}\ \bibnamefont {Ferrari}},\ }\href@noop {}
  {\bibfield  {journal} {\bibinfo  {journal} {Nature Photonics}\ }\textbf
  {\bibinfo {volume} {4}},\ \bibinfo {pages} {611} (\bibinfo {year}
  {2010})}\BibitemShut {NoStop}%
\bibitem [{\citenamefont {Castro~Neto}\ \emph {et~al.}(2009)\citenamefont
  {Castro~Neto}, \citenamefont {Guinea}, \citenamefont {Peres}, \citenamefont
  {Novoselov},\ and\ \citenamefont {Geim}}]{graphene_electric}%
  \BibitemOpen
  \bibfield  {author} {\bibinfo {author} {\bibfnamefont {A.~H.}\ \bibnamefont
  {Castro~Neto}}, \bibinfo {author} {\bibfnamefont {F.}~\bibnamefont {Guinea}},
  \bibinfo {author} {\bibfnamefont {N.~M.~R.}\ \bibnamefont {Peres}}, \bibinfo
  {author} {\bibfnamefont {K.~S.}\ \bibnamefont {Novoselov}}, \ and\ \bibinfo
  {author} {\bibfnamefont {A.~K.}\ \bibnamefont {Geim}},\ }\href@noop {}
  {\bibfield  {journal} {\bibinfo  {journal} {Reviews of Modern Physics}\
  }\textbf {\bibinfo {volume} {81}},\ \bibinfo {pages} {109} (\bibinfo {year}
  {2009})}\BibitemShut {NoStop}%
\bibitem [{\citenamefont {Singh}\ \emph {et~al.}(2011)\citenamefont {Singh},
  \citenamefont {Joung}, \citenamefont {Zhai}, \citenamefont {Das},
  \citenamefont {Khondaker},\ and\ \citenamefont {Seal}}]{graphene_structure}%
  \BibitemOpen
  \bibfield  {author} {\bibinfo {author} {\bibfnamefont {V.}~\bibnamefont
  {Singh}}, \bibinfo {author} {\bibfnamefont {D.}~\bibnamefont {Joung}},
  \bibinfo {author} {\bibfnamefont {L.}~\bibnamefont {Zhai}}, \bibinfo {author}
  {\bibfnamefont {S.}~\bibnamefont {Das}}, \bibinfo {author} {\bibfnamefont
  {S.~I.}\ \bibnamefont {Khondaker}}, \ and\ \bibinfo {author} {\bibfnamefont
  {S.}~\bibnamefont {Seal}},\ }\href@noop {} {\bibfield  {journal} {\bibinfo
  {journal} {Progress in Materials Science}\ }\textbf {\bibinfo {volume}
  {56}},\ \bibinfo {pages} {1178} (\bibinfo {year} {2011})}\BibitemShut
  {NoStop}%
\bibitem [{\citenamefont {Wang}\ \emph {et~al.}(2012)\citenamefont {Wang},
  \citenamefont {Kalantar-Zadeh}, \citenamefont {Kis}, \citenamefont
  {Coleman},\ and\ \citenamefont {Strano}}]{2D_tmds}%
  \BibitemOpen
  \bibfield  {author} {\bibinfo {author} {\bibfnamefont {Q.~H.}\ \bibnamefont
  {Wang}}, \bibinfo {author} {\bibfnamefont {K.}~\bibnamefont
  {Kalantar-Zadeh}}, \bibinfo {author} {\bibfnamefont {A.}~\bibnamefont {Kis}},
  \bibinfo {author} {\bibfnamefont {J.~N.}\ \bibnamefont {Coleman}}, \ and\
  \bibinfo {author} {\bibfnamefont {M.~S.}\ \bibnamefont {Strano}},\
  }\href@noop {} {\bibfield  {journal} {\bibinfo  {journal} {Nature
  Nanotechnology}\ }\textbf {\bibinfo {volume} {7}},\ \bibinfo {pages} {699}
  (\bibinfo {year} {2012})}\BibitemShut {NoStop}%
\bibitem [{\citenamefont {Mak}\ \emph {et~al.}(2010)\citenamefont {Mak},
  \citenamefont {Lee}, \citenamefont {Hone}, \citenamefont {Shan},\ and\
  \citenamefont {Heinz}}]{tmd_direct_gap}%
  \BibitemOpen
  \bibfield  {author} {\bibinfo {author} {\bibfnamefont {K.~F.}\ \bibnamefont
  {Mak}}, \bibinfo {author} {\bibfnamefont {C.}~\bibnamefont {Lee}}, \bibinfo
  {author} {\bibfnamefont {J.}~\bibnamefont {Hone}}, \bibinfo {author}
  {\bibfnamefont {J.}~\bibnamefont {Shan}}, \ and\ \bibinfo {author}
  {\bibfnamefont {T.~F.}\ \bibnamefont {Heinz}},\ }\href@noop {} {\bibfield
  {journal} {\bibinfo  {journal} {Physical Review Letters}\ }\textbf {\bibinfo
  {volume} {105}} (\bibinfo {year} {2010})}\BibitemShut {NoStop}%
\bibitem [{\citenamefont {Mak}\ and\ \citenamefont {Shan}(2016)}]{opto_2D}%
  \BibitemOpen
  \bibfield  {author} {\bibinfo {author} {\bibfnamefont {K.~F.}\ \bibnamefont
  {Mak}}\ and\ \bibinfo {author} {\bibfnamefont {J.}~\bibnamefont {Shan}},\
  }\href@noop {} {\bibfield  {journal} {\bibinfo  {journal} {Nature Photonics}\
  }\textbf {\bibinfo {volume} {10}},\ \bibinfo {pages} {216} (\bibinfo {year}
  {2016})}\BibitemShut {NoStop}%
\bibitem [{\citenamefont {Novoselov}\ \emph {et~al.}(2005)\citenamefont
  {Novoselov}, \citenamefont {Jiang}, \citenamefont {Schedin}, \citenamefont
  {Booth}, \citenamefont {Khotkevich}, \citenamefont {Morozov},\ and\
  \citenamefont {Geim}}]{exciton_low_binding}%
  \BibitemOpen
  \bibfield  {author} {\bibinfo {author} {\bibfnamefont {K.}~\bibnamefont
  {Novoselov}}, \bibinfo {author} {\bibfnamefont {D.}~\bibnamefont {Jiang}},
  \bibinfo {author} {\bibfnamefont {F.}~\bibnamefont {Schedin}}, \bibinfo
  {author} {\bibfnamefont {T.}~\bibnamefont {Booth}}, \bibinfo {author}
  {\bibfnamefont {V.}~\bibnamefont {Khotkevich}}, \bibinfo {author}
  {\bibfnamefont {S.}~\bibnamefont {Morozov}}, \ and\ \bibinfo {author}
  {\bibfnamefont {A.}~\bibnamefont {Geim}},\ }\href@noop {} {\bibfield
  {journal} {\bibinfo  {journal} {Proceedings of The National Academy of
  Sciences of The United States of America}\ }\textbf {\bibinfo {volume}
  {102}},\ \bibinfo {pages} {10451} (\bibinfo {year} {2005})}\BibitemShut
  {NoStop}%
\bibitem [{\citenamefont {Hanbicki}\ \emph {et~al.}(2015)\citenamefont
  {Hanbicki}, \citenamefont {Currie}, \citenamefont {Kioseoglou}, \citenamefont
  {Friedman},\ and\ \citenamefont {Jonker}}]{hanbicki_2015}%
  \BibitemOpen
  \bibfield  {author} {\bibinfo {author} {\bibfnamefont {A.~T.}\ \bibnamefont
  {Hanbicki}}, \bibinfo {author} {\bibfnamefont {M.}~\bibnamefont {Currie}},
  \bibinfo {author} {\bibfnamefont {G.}~\bibnamefont {Kioseoglou}}, \bibinfo
  {author} {\bibfnamefont {A.~L.}\ \bibnamefont {Friedman}}, \ and\ \bibinfo
  {author} {\bibfnamefont {B.~T.}\ \bibnamefont {Jonker}},\ }\href@noop {}
  {\bibfield  {journal} {\bibinfo  {journal} {Solid State Communications}\
  }\textbf {\bibinfo {volume} {203}},\ \bibinfo {pages} {16} (\bibinfo {year}
  {2015})}\BibitemShut {NoStop}%
\bibitem [{\citenamefont {Knox}(1963)}]{exciton_theory}%
  \BibitemOpen
  \bibfield  {author} {\bibinfo {author} {\bibfnamefont {R.~S.}\ \bibnamefont
  {Knox}},\ }\href@noop {} {\  (\bibinfo {year} {Academic, 1963})}\BibitemShut
  {NoStop}%
\bibitem [{\citenamefont {Nozik}\ \emph {et~al.}(2010)\citenamefont {Nozik},
  \citenamefont {Beard}, \citenamefont {Luther}, \citenamefont {Law},
  \citenamefont {Ellingson},\ and\ \citenamefont {Johnson}}]{exciton_app1}%
  \BibitemOpen
  \bibfield  {author} {\bibinfo {author} {\bibfnamefont {A.~J.}\ \bibnamefont
  {Nozik}}, \bibinfo {author} {\bibfnamefont {M.~C.}\ \bibnamefont {Beard}},
  \bibinfo {author} {\bibfnamefont {J.~M.}\ \bibnamefont {Luther}}, \bibinfo
  {author} {\bibfnamefont {M.}~\bibnamefont {Law}}, \bibinfo {author}
  {\bibfnamefont {R.~J.}\ \bibnamefont {Ellingson}}, \ and\ \bibinfo {author}
  {\bibfnamefont {J.~C.}\ \bibnamefont {Johnson}},\ }\href@noop {} {\bibfield
  {journal} {\bibinfo  {journal} {Chemical Reviews}\ }\textbf {\bibinfo
  {volume} {110}},\ \bibinfo {pages} {6873} (\bibinfo {year}
  {2010})}\BibitemShut {NoStop}%
\bibitem [{\citenamefont {Zhao}\ \emph {et~al.}(2012)\citenamefont {Zhao},
  \citenamefont {Yu}, \citenamefont {Shan}, \citenamefont {Wang}, \citenamefont
  {Xu},\ and\ \citenamefont {Chen}}]{exciton_app2}%
  \BibitemOpen
  \bibfield  {author} {\bibinfo {author} {\bibfnamefont {W.-W.}\ \bibnamefont
  {Zhao}}, \bibinfo {author} {\bibfnamefont {P.-P.}\ \bibnamefont {Yu}},
  \bibinfo {author} {\bibfnamefont {Y.}~\bibnamefont {Shan}}, \bibinfo {author}
  {\bibfnamefont {J.}~\bibnamefont {Wang}}, \bibinfo {author} {\bibfnamefont
  {J.-J.}\ \bibnamefont {Xu}}, \ and\ \bibinfo {author} {\bibfnamefont {H.-Y.}\
  \bibnamefont {Chen}},\ }\href@noop {} {\bibfield  {journal} {\bibinfo
  {journal} {Analytical Chemistry}\ }\textbf {\bibinfo {volume} {84}},\
  \bibinfo {pages} {5892} (\bibinfo {year} {2012})}\BibitemShut {NoStop}%
\bibitem [{\citenamefont {Tseng}\ \emph {et~al.}(2016)\citenamefont {Tseng},
  \citenamefont {Simsek},\ and\ \citenamefont {Gunlycke}}]{simsek_storage}%
  \BibitemOpen
  \bibfield  {author} {\bibinfo {author} {\bibfnamefont {F.}~\bibnamefont
  {Tseng}}, \bibinfo {author} {\bibfnamefont {E.}~\bibnamefont {Simsek}}, \
  and\ \bibinfo {author} {\bibfnamefont {D.}~\bibnamefont {Gunlycke}},\
  }\href@noop {} {\bibfield  {journal} {\bibinfo  {journal} {Journal of
  Physics-Condensed Matter}\ }\textbf {\bibinfo {volume} {28}} (\bibinfo {year}
  {2016})}\BibitemShut {NoStop}%
\bibitem [{\citenamefont {Ye}\ \emph {et~al.}(2014)\citenamefont {Ye},
  \citenamefont {Cao}, \citenamefont {O'Brien}, \citenamefont {Zhu},
  \citenamefont {Yin}, \citenamefont {Wang}, \citenamefont {Louie},\ and\
  \citenamefont {Zhang}}]{cao_2014}%
  \BibitemOpen
  \bibfield  {author} {\bibinfo {author} {\bibfnamefont {Z.}~\bibnamefont
  {Ye}}, \bibinfo {author} {\bibfnamefont {T.}~\bibnamefont {Cao}}, \bibinfo
  {author} {\bibfnamefont {K.}~\bibnamefont {O'Brien}}, \bibinfo {author}
  {\bibfnamefont {H.}~\bibnamefont {Zhu}}, \bibinfo {author} {\bibfnamefont
  {X.}~\bibnamefont {Yin}}, \bibinfo {author} {\bibfnamefont {Y.}~\bibnamefont
  {Wang}}, \bibinfo {author} {\bibfnamefont {S.~G.}\ \bibnamefont {Louie}}, \
  and\ \bibinfo {author} {\bibfnamefont {X.}~\bibnamefont {Zhang}},\
  }\href@noop {} {\bibfield  {journal} {\bibinfo  {journal} {Nature}\ }\textbf
  {\bibinfo {volume} {513}},\ \bibinfo {pages} {214} (\bibinfo {year}
  {2014})}\BibitemShut {NoStop}%
\bibitem [{\citenamefont {Chernikov}\ \emph {et~al.}(2014)\citenamefont
  {Chernikov}, \citenamefont {Berkelbach}, \citenamefont {Hill}, \citenamefont
  {Rigosi}, \citenamefont {Li}, \citenamefont {Aslan}, \citenamefont
  {Reichman}, \citenamefont {Hybertsen},\ and\ \citenamefont
  {Heinz}}]{chernikov_2014}%
  \BibitemOpen
  \bibfield  {author} {\bibinfo {author} {\bibfnamefont {A.}~\bibnamefont
  {Chernikov}}, \bibinfo {author} {\bibfnamefont {T.~C.}\ \bibnamefont
  {Berkelbach}}, \bibinfo {author} {\bibfnamefont {H.~M.}\ \bibnamefont
  {Hill}}, \bibinfo {author} {\bibfnamefont {A.}~\bibnamefont {Rigosi}},
  \bibinfo {author} {\bibfnamefont {Y.}~\bibnamefont {Li}}, \bibinfo {author}
  {\bibfnamefont {O.~B.}\ \bibnamefont {Aslan}}, \bibinfo {author}
  {\bibfnamefont {D.~R.}\ \bibnamefont {Reichman}}, \bibinfo {author}
  {\bibfnamefont {M.~S.}\ \bibnamefont {Hybertsen}}, \ and\ \bibinfo {author}
  {\bibfnamefont {T.~F.}\ \bibnamefont {Heinz}},\ }\href@noop {} {\bibfield
  {journal} {\bibinfo  {journal} {Physical Review Letters}\ }\textbf {\bibinfo
  {volume} {113}} (\bibinfo {year} {2014})}\BibitemShut {NoStop}%
\bibitem [{\citenamefont {Ramasubramaniam}(2012)}]{exciton_theory1}%
  \BibitemOpen
  \bibfield  {author} {\bibinfo {author} {\bibfnamefont {A.}~\bibnamefont
  {Ramasubramaniam}},\ }\href@noop {} {\bibfield  {journal} {\bibinfo
  {journal} {Physical Review B}\ }\textbf {\bibinfo {volume} {86}} (\bibinfo
  {year} {2012})}\BibitemShut {NoStop}%
\bibitem [{\citenamefont {Shi}\ \emph {et~al.}(2013)\citenamefont {Shi},
  \citenamefont {Pan}, \citenamefont {Zhang},\ and\ \citenamefont
  {Yakobson}}]{exciton_theory2}%
  \BibitemOpen
  \bibfield  {author} {\bibinfo {author} {\bibfnamefont {H.}~\bibnamefont
  {Shi}}, \bibinfo {author} {\bibfnamefont {H.}~\bibnamefont {Pan}}, \bibinfo
  {author} {\bibfnamefont {Y.-W.}\ \bibnamefont {Zhang}}, \ and\ \bibinfo
  {author} {\bibfnamefont {B.~I.}\ \bibnamefont {Yakobson}},\ }\href@noop {}
  {\bibfield  {journal} {\bibinfo  {journal} {Physical Review B}\ }\textbf
  {\bibinfo {volume} {87}} (\bibinfo {year} {2013})}\BibitemShut {NoStop}%
\bibitem [{\citenamefont {Berkelbach}\ \emph {et~al.}(2013)\citenamefont
  {Berkelbach}, \citenamefont {Hybertsen},\ and\ \citenamefont
  {Reichman}}]{exciton_theory3}%
  \BibitemOpen
  \bibfield  {author} {\bibinfo {author} {\bibfnamefont {T.~C.}\ \bibnamefont
  {Berkelbach}}, \bibinfo {author} {\bibfnamefont {M.~S.}\ \bibnamefont
  {Hybertsen}}, \ and\ \bibinfo {author} {\bibfnamefont {D.~R.}\ \bibnamefont
  {Reichman}},\ }\href@noop {} {\bibfield  {journal} {\bibinfo  {journal}
  {Physical Review B}\ }\textbf {\bibinfo {volume} {88}} (\bibinfo {year}
  {2013})}\BibitemShut {NoStop}%
\bibitem [{\citenamefont {Komsa}\ and\ \citenamefont
  {Krasheninnikov}(2012)}]{binding_substrate}%
  \BibitemOpen
  \bibfield  {author} {\bibinfo {author} {\bibfnamefont {H.-P.}\ \bibnamefont
  {Komsa}}\ and\ \bibinfo {author} {\bibfnamefont {A.~V.}\ \bibnamefont
  {Krasheninnikov}},\ }\href@noop {} {\bibfield  {journal} {\bibinfo  {journal}
  {Physical Review B}\ }\textbf {\bibinfo {volume} {86}},\ \bibinfo {pages}
  {241201} (\bibinfo {year} {2012})}\BibitemShut {NoStop}%
\bibitem [{\citenamefont {Gunlycke}\ and\ \citenamefont
  {Tseng}(2016)}]{triangular_exciton}%
  \BibitemOpen
  \bibfield  {author} {\bibinfo {author} {\bibfnamefont {D.}~\bibnamefont
  {Gunlycke}}\ and\ \bibinfo {author} {\bibfnamefont {F.}~\bibnamefont
  {Tseng}},\ }\href@noop {} {\bibfield  {journal} {\bibinfo  {journal}
  {Physical Chemistry Chemical Physics}\ }\textbf {\bibinfo {volume} {18}},\
  \bibinfo {pages} {8579} (\bibinfo {year} {2016})}\BibitemShut {NoStop}%
\bibitem [{\citenamefont {Whitney}\ and\ \citenamefont
  {Crater}(2014)}]{baryon_spectrum}%
  \BibitemOpen
  \bibfield  {author} {\bibinfo {author} {\bibfnamefont {J.~F.}\ \bibnamefont
  {Whitney}}\ and\ \bibinfo {author} {\bibfnamefont {H.~W.}\ \bibnamefont
  {Crater}},\ }\href@noop {} {\bibfield  {journal} {\bibinfo  {journal} {Phys.
  Rev. D}\ }\textbf {\bibinfo {volume} {89}},\ \bibinfo {pages} {014023}
  (\bibinfo {year} {2014})}\BibitemShut {NoStop}%
\bibitem [{\citenamefont {Van~Alstine}\ and\ \citenamefont
  {Crater}(1997)}]{tale_3_equations}%
  \BibitemOpen
  \bibfield  {author} {\bibinfo {author} {\bibfnamefont {P.}~\bibnamefont
  {Van~Alstine}}\ and\ \bibinfo {author} {\bibfnamefont {H.~W.}\ \bibnamefont
  {Crater}},\ }\href@noop {} {\bibfield  {journal} {\bibinfo  {journal}
  {Foundations of Physics}\ }\textbf {\bibinfo {volume} {27}},\ \bibinfo
  {pages} {67} (\bibinfo {year} {1997})}\BibitemShut {NoStop}%
\bibitem [{\citenamefont {Breit}(1929)}]{breit_1929}%
  \BibitemOpen
  \bibfield  {author} {\bibinfo {author} {\bibfnamefont {G.}~\bibnamefont
  {Breit}},\ }\href {\doibase 10.1103/PhysRev.34.553} {\bibfield  {journal}
  {\bibinfo  {journal} {Phys. Rev.}\ }\textbf {\bibinfo {volume} {34}},\
  \bibinfo {pages} {553} (\bibinfo {year} {1929})}\BibitemShut {NoStop}%
\bibitem [{\citenamefont {Salpeter}\ and\ \citenamefont
  {Bethe}(1951)}]{bethe_salpeter_1951}%
  \BibitemOpen
  \bibfield  {author} {\bibinfo {author} {\bibfnamefont {E.~E.}\ \bibnamefont
  {Salpeter}}\ and\ \bibinfo {author} {\bibfnamefont {H.~A.}\ \bibnamefont
  {Bethe}},\ }\href@noop {} {\bibfield  {journal} {\bibinfo  {journal} {Phys.
  Rev.}\ }\textbf {\bibinfo {volume} {84}},\ \bibinfo {pages} {1232} (\bibinfo
  {year} {1951})}\BibitemShut {NoStop}%
\bibitem [{\citenamefont {Kemmer}(1937)}]{kemmer_1937}%
  \BibitemOpen
  \bibfield  {author} {\bibinfo {author} {\bibfnamefont {N.}~\bibnamefont
  {Kemmer}},\ }\href@noop {} {\bibfield  {journal} {\bibinfo  {journal} {Phys.
  Rev.}\ }\textbf {\bibinfo {volume} {52}},\ \bibinfo {pages} {906} (\bibinfo
  {year} {1937})}\BibitemShut {NoStop}%
\bibitem [{\citenamefont {Fermi}\ and\ \citenamefont
  {Yang}(1949)}]{fermi-yang}%
  \BibitemOpen
  \bibfield  {author} {\bibinfo {author} {\bibfnamefont {E.}~\bibnamefont
  {Fermi}}\ and\ \bibinfo {author} {\bibfnamefont {C.~N.}\ \bibnamefont
  {Yang}},\ }\href@noop {} {\bibfield  {journal} {\bibinfo  {journal} {Phys.
  Rev.}\ }\textbf {\bibinfo {volume} {76}},\ \bibinfo {pages} {1739} (\bibinfo
  {year} {1949})}\BibitemShut {NoStop}%
\bibitem [{\citenamefont {Barut}\ and\ \citenamefont
  {Komy}(1985)}]{barut_derivation_1985}%
  \BibitemOpen
  \bibfield  {author} {\bibinfo {author} {\bibfnamefont {A.~O.}\ \bibnamefont
  {Barut}}\ and\ \bibinfo {author} {\bibfnamefont {S.}~\bibnamefont {Komy}},\
  }\href@noop {} {\bibfield  {journal} {\bibinfo  {journal} {Fortschritte der
  Physik/Progress of Physics}\ }\textbf {\bibinfo {volume} {33}},\ \bibinfo
  {pages} {309} (\bibinfo {year} {1985})}\BibitemShut {NoStop}%
\bibitem [{\citenamefont {Barut}\ and\ \citenamefont
  {Unal}(1985)}]{barut_radial_1985}%
  \BibitemOpen
  \bibfield  {author} {\bibinfo {author} {\bibfnamefont {A.~O.}\ \bibnamefont
  {Barut}}\ and\ \bibinfo {author} {\bibfnamefont {N.}~\bibnamefont {Unal}},\
  }\href@noop {} {\bibfield  {journal} {\bibinfo  {journal} {Fortschritte der
  Physik/Progress of Physics}\ }\textbf {\bibinfo {volume} {33}},\ \bibinfo
  {pages} {319} (\bibinfo {year} {1985})}\BibitemShut {NoStop}%
\bibitem [{\citenamefont {Crater}\ and\ \citenamefont
  {Van~Alstine}(1988)}]{meson_spectroscopy}%
  \BibitemOpen
  \bibfield  {author} {\bibinfo {author} {\bibfnamefont {H.~W.}\ \bibnamefont
  {Crater}}\ and\ \bibinfo {author} {\bibfnamefont {P.}~\bibnamefont
  {Van~Alstine}},\ }\href@noop {} {\bibfield  {journal} {\bibinfo  {journal}
  {Phys. Rev. D}\ }\textbf {\bibinfo {volume} {37}},\ \bibinfo {pages} {1982}
  (\bibinfo {year} {1988})}\BibitemShut {NoStop}%
\bibitem [{\citenamefont {Aydin}\ and\ \citenamefont
  {Yilmazer}(1988)}]{relativistic_two_fermion_problem}%
  \BibitemOpen
  \bibfield  {author} {\bibinfo {author} {\bibfnamefont {Z.~Z.}\ \bibnamefont
  {Aydin}}\ and\ \bibinfo {author} {\bibfnamefont {A.~U.}\ \bibnamefont
  {Yilmazer}},\ }\href@noop {} {\bibfield  {journal} {\bibinfo  {journal}
  {Journal of Physics G: Nuclear Physics}\ }\textbf {\bibinfo {volume} {14}},\
  \bibinfo {pages} {1345} (\bibinfo {year} {1988})}\BibitemShut {NoStop}%
\bibitem [{\citenamefont {Berman}\ \emph {et~al.}(2013)\citenamefont {Berman},
  \citenamefont {Kezerashvili},\ and\ \citenamefont {Ziegler}}]{oleg-klaus}%
  \BibitemOpen
  \bibfield  {author} {\bibinfo {author} {\bibfnamefont {O.~L.}\ \bibnamefont
  {Berman}}, \bibinfo {author} {\bibfnamefont {R.~Y.}\ \bibnamefont
  {Kezerashvili}}, \ and\ \bibinfo {author} {\bibfnamefont {K.}~\bibnamefont
  {Ziegler}},\ }\href@noop {} {\bibfield  {journal} {\bibinfo  {journal} {Phys.
  Rev. A}\ }\textbf {\bibinfo {volume} {87}},\ \bibinfo {pages} {042513}
  (\bibinfo {year} {2013})}\BibitemShut {NoStop}%
\bibitem [{\citenamefont {Dong}\ and\ \citenamefont
  {Ma}(2003)}]{dong_exact_2003}%
  \BibitemOpen
  \bibfield  {author} {\bibinfo {author} {\bibfnamefont {S.-H.}\ \bibnamefont
  {Dong}}\ and\ \bibinfo {author} {\bibfnamefont {Z.-Q.}\ \bibnamefont {Ma}},\
  }\href@noop {} {\bibfield  {journal} {\bibinfo  {journal} {Physics Letters
  A}\ }\textbf {\bibinfo {volume} {312}},\ \bibinfo {pages} {78} (\bibinfo
  {year} {2003})}\BibitemShut {NoStop}%
\bibitem [{\citenamefont {Witten}(1988)}]{witten_1988}%
  \BibitemOpen
  \bibfield  {author} {\bibinfo {author} {\bibfnamefont {E.}~\bibnamefont
  {Witten}},\ }\href@noop {} {\bibfield  {journal} {\bibinfo  {journal}
  {Nuclear Physics B}\ }\textbf {\bibinfo {volume} {311}},\ \bibinfo {pages}
  {46} (\bibinfo {year} {1988})}\BibitemShut {NoStop}%
\end{thebibliography}

\end{document}